# Engineering Andreev Bound States for Thermal Sensing in Proximity Josephson Junctions

Woochan Jung[1], Ethan G Arnault[2], Bevin Huang[3], Jinho Park[1,4], Seong Jang[1], Kenji Watanabe[5], Takashi Taniguchi[6], Dirk Englund[7], Kin Chung Fong[8,9,10,*], Gil-Ho Lee[1,*]

[1]Physics Department, Pohang University of Science and Technology, Pohang 37673, South Korea
[2]Research Laboratory of Electronics, Massachusetts Institute of Technology, Cambridge, MA 02139, USA
[3]Intelligence Community Postdoctoral Research Fellowship Program, Massachusetts Institute of Technology, Cambridge, MA 02139, USA
[4]Department of Mechanical Engineering, Columbia University, New York, NY 10027, USA
[5]Research Center for Functional Materials, National Institute for Materials Science, Tsukuba, Japan
[6]International Center for Materials Nanoarchitectonics, National Institute for Materials Science, Tsukuba, Japan
[7]Department of Electrical Engineering and Computer Science, Massachusetts Institute of Technology, Cambridge, MA 02139, USA
[8]Department of Electrical and Computer Engineering, Northeastern University, Boston, MA 02115, USA
[9]Department of Physics, Northeastern University, Boston, MA 02115, USA
[10]Quantum Materials and Sensing Institute, Burlington, MA 01803, USA

*Corresponding authors: k.fong@northeastern.edu (K.C.F.), lghman@postech.ac.kr (G.-H.L.)

**ABSTRACT**. The thermal response of proximity Josephson junctions (JJs) is governed by the temperature ($T$)-dependent occupation of Andreev bound states (ABS), making them promising candidates for sensitive thermal detection. In this study, we systematically engineer ABS to enhance the thermal sensitivity of critical current ($I_c$) of proximity JJs, quantified as $|dI_c/dT|$ for threshold readout scheme and $|dI_c/dT \cdot I_c^{-1}|$ for inductive readout scheme. Using a gate-tunable graphene-based JJ platform, we explore the impact of key parameters—including channel length, transparency, carrier density, and superconducting material—on the thermal response. Our results reveal that the proximity-induced superconducting gap plays a crucial role in optimizing thermal sensitivity. Notably, we see a maximum $|dI_c/dT \cdot I_c^{-1}|$ value of 0.6 K$^{-1}$ at low temperatures with titanium-based graphene JJs. By demonstrating a systematic approach to engineering ABS in proximity JJs, this work establishes a versatile framework for optimizing thermal sensors and advancing the study of ABS-mediated transport.

## I. INTRODUCTION

As the mainstay of both classical and quantum superconducting technologies, the Josephson junction (JJ) plays a vital role in superconducting qubits [1], quantum-limited amplifiers [2, 3], quantum sensors [4–7], and other fundamental scientific applications [8–13]. More recently, JJs have gained attention as highly effective thermal sensors [14–27], leveraging their fast-switching behavior for exceptional detection speed and low-heat capacity for large thermal responses. For instance, nanosecond thermometry with nanobridge-type JJs have already been demonstrated [21]. Moreover, Josephson junction bolometers have been proposed [15] and demonstrated a noise equivalent power (NEP) of sub-aW/$\sqrt{\text{Hz}}$ [18, 22, 23]. JJs have also been employed in calorimetric sensor designs [15, 20], including single-photon detection (SPD) capabilities [27, 28].

In a proximity JJ, which consists of a superconductor/normal-metal/superconductor structure, thermal sensitivity arises from the temperature-dependent behavior of Andreev bound states (ABS) in which the energy scale is determined by the proximity-induced superconducting gap ($\Delta^*$) [29]. As temperature fluctuations alter the occupation of these ABS, the critical current of the JJ varies accordingly. Unlike tunnel JJs, in which the critical current at low temperatures exhibits weaker temperature dependence according to the Ambegaokar–Baratoff relation [30], proximity JJs remain highly sensitive in this regime owing to the presence of ABS with energy scales smaller than the proximity-induced superconducting gap in the normal-metal. Previous studies on superconducting gap engineering in proximity JJs have primarily addressed topics such as quasiparticle poisoning [31–35], parity flips in ABS spectra [34–36], or overheating of Majorana bound states [37], rather than focusing on the application of JJs for thermal sensing.

A deeper understanding of how the proximity-induced superconducting gap affects the temperature dependence of proximity JJs is crucial for optimizing their thermal sensitivity and enhancing proximity JJ-based thermal sensor performance. This approach parallels recent advancements in gap engineering for tunnel JJs that aim to minimize decoherence by suppressing quasiparticle tunneling [38–43].

To confine our experimental efforts, we will focus on graphene JJ (GJJs), but the physics of how the proximity-induced superconducting gap in graphene influences the behavior of ABS is general and should be applicable to other proximity junctions and sensors [14–17, 20, 22–25, 27]. In addition to its gate tunability, we chose to study GJJs because of its promising thermal properties [22, 23, 25, 27] for detecting low-energy photons. Owing to its single-atom thickness and Dirac-like electron spectrum, graphene exhibits an ultralow electronic specific heat capacity ($\sim 1\ k_B/\mu m^2$) near the charge neutrality point (CNP) [44]. Consequently, even a small energy absorption can lead to a major increase in electron temperature ($T_e$). Furthermore, graphene's rapid electron–electron interactions (~10 fs) [45] enable efficient thermalization, while its weak electron–phonon coupling [46, 47] retains the thermal energy for sensing.

Graphene's zero-bandgap nature allows broadband photon absorption and highly transparent superconducting contacts, making it a promising material for overcoming the bandwidth limitations of conventional sensors [48–52]. A graphene-based thermal sensor can extend the detection range to GHz frequencies without requiring tuning [53]. Recent experiments have demonstrated the viability of GJJs as bolometers [22, 23], achieving sufficient NEP levels to detect single 32 GHz photons [22]. Additionally, SPD has been successfully realized in the near-infrared range [25, 27]. However, a systematic study focused on leveraging these capabilities for thermal sensor enhancement remains lacking.

This work aims to engineer the thermal response of ABS to optimize the performance of proximity JJs as thermal sensors. To achieve this, we investigate the temperature dependence of proximity JJ across various device parameters—channel length, channel width, back-gate voltage, contact transparency, and the choice of superconducting materials.

## II. Readout schemes for thermal sensing

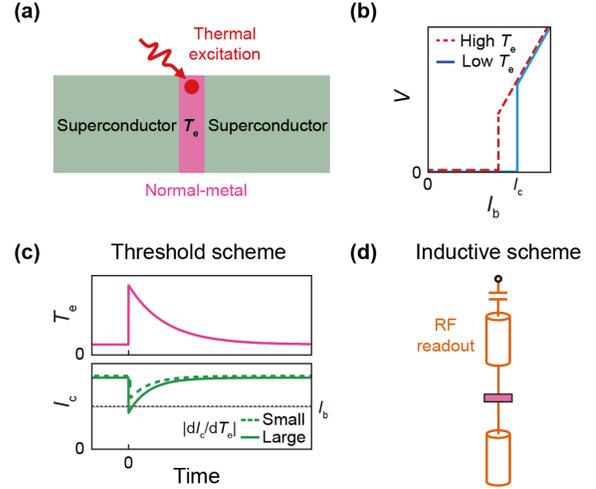

FIG 1. Schematics of thermal sensing. (a) Device schematics of proximity Josephson junction (JJ)-based thermal sensor. (b) Bias current ($I_b$)–junction voltage ($V$) characteristics yield a critical current ($I_c$) above which the voltage switches abruptly from a supercurrent to a resistive state. (c) Upper panel shows the temporal profile of the electron temperature of normal-metal ($T_e$). Lower panel shows the temporal profile of $I_c$ for both cases of small (dashed line) and large (solid line) $|dI_c/dT_e|$ values with fixed $I_b$ (horizontal dotted line). (d) Schematics of the Josephson inductance readout setup with readout resonator (orange) integrated with normal-metal part (pink rectangle) of proximity JJ.

We study the thermal response of ABS through $|dI_c/dT_e|$ and $|dI_c/dT_e \cdot I_c^{-1}|$ as these are the most relevant parameters governing the performance of proximity JJ-based thermal sensors [15]. Figure 1a illustrates the abstracted schematic of the proximity JJ-based thermal sensor. In this design, thermal excitations are absorbed by the weak link made of normal-metal and alter the Josephson coupling strength. The operation of the detectors can be divided into two categories—DC threshold and AC inductive schemes, which could be optimized using $|dI_c/dT_e|$ and $|dI_c/dT_e \cdot I_c^{-1}|$, respectively. Both approaches exploit the temperature dependence of $I_c$ (Fig. 1b) to generate an electrical response upon thermal excitation.

The DC threshold scheme (Fig. 1c) is analogous to that used in superconducting nanowire single-photon detectors [50]. With a bias current ($I_b$) near $I_c$, the detector can transition from the supercurrent to a resistive state triggered by a thermal excitation. A large $|dI_c/dT_e|$ value ensures that even a small thermal excitation lowers $I_c$ below $I_b$, producing a detectable voltage signal. However, if $|dI_c/dT_e|$ is

small, $I_c$ remains above $I_b$, preventing the switching event. Thus, $|dI_c/dT_e|$ is a crucial figure of merit for this detection scheme [20]. Increasing $I_b$ close to $I_c$ enhances detection efficiency but also increases the dark count rate.

Similar to microwave kinetic inductance detectors [48], in the inductive scheme (Fig. 1d), thermal excitations modify $T_e$—detected through a shift in the resonance frequency, $\delta f_r$, of an LC circuit composed of the proximity JJ. As the Josephson inductance $L_J$ is inversely proportional to $I_c$, any temperature-induced change in $I_c$ directly affects $L_J$. This approach allows high-speed measurements on the nanosecond scale. Additionally, because the junction remains in the supercurrent state, the device could avoid the extra DC Joule heating when $I_b$ exceeds $I_c$ in the threshold scheme and hence minimize the dead time. When $L_J \ll L_{res}$, where $L_{res}$ is an intrinsic inductance of the readout resonator, the shift in Josephson inductance $\delta L_J$ due to the change of $T_e$ in the perturbation regime can be expressed as,

$$-2\frac{\delta f_r}{f_r} = \frac{\delta L_J}{L_J + L_{res}} \propto |dI_c/dT_e \cdot I_c^{-1}|. \tag{1}$$

The above equation indicates that $|dI_c/dT_e \cdot I_c^{-1}|$ is the key metric governing the readout sensitivity in the inductive readout scheme [54].

### III. RESULTS AND DISCUSSION

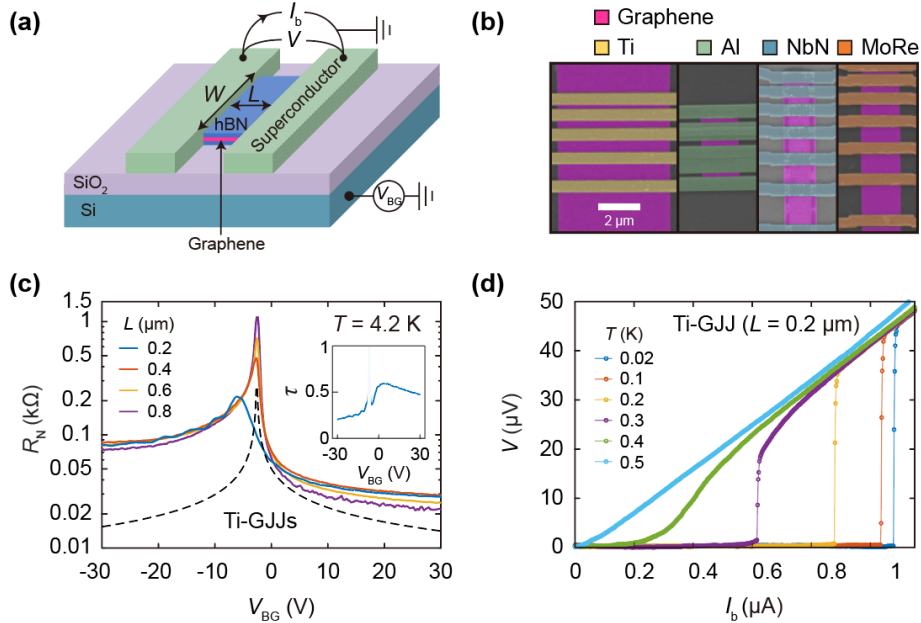

FIG 2. Direct current characteristics of GJJs. (a) Schematic of GJJ with graphene (pink) encapsulated by hexagonal boron nitride (hBN) layers (blue) and contacted by superconducting electrodes (green). The schematic depicts the channel's width $W$ and channel's length $L$ of GJJ, back-gate voltage $V_{BG}$, bias current $I_b$, and junction voltage V. (b) False-colored scanning electron microscopy images of GJJs fabricated with various superconducting materials, including Ti, Al, NbN, and MoRe alloy. (c) Gate-voltage dependence of the normal-state junction resistance $R_N$ for Ti-GJJs with difference $L$ at the temperature $T = 4.2$ K, higher than the superconducting transition temperature of Ti. The dashed line represents $R_N$ expected in the ballistic limit with perfect contact transparency ($\tau = 1$). The inset shows the variation of the contact transparency $\tau$ as a function of $V_{BG}$ for a Ti-GJJ with $L = 0.2$ μm. (d) $I_b$–V characteristic of Ti-GJJ with $L = 0.2$ μm at various temperatures measured by sweeping $I_b$ from zero to positive values.

Taking advantage of its gate-tunability and versatile connectivity with various superconducting materials to form a proximity JJs, we use GJJs as a model platform to study the thermal dependence of ABS and optimize their thermal detector performance. We fabricated and measured GJJs with various junction parameters, including the channel's length $(L)$, channel's width $(W)$, back-gate voltage $(V_{BG})$, contact transparency $(\tau)$, and the types of superconducting materials. The graphene was encapsulated by top and bottom hexagonal boron nitride (hBN) insulating layers to minimize impurity scattering within the graphene, and superconducting electrodes were contacted at the edge of the graphene, as depicted in Fig. 2a (See Methods in Supplementary Information for details). Superconducting materials for GJJs include titanium (Ti), aluminum (Al), niobium nitride (NbN), and molybdenum–rhenium alloy (MoRe), as shown in Fig. 2b (see Table I and Supplementary Fig. 1 for device information).

We begin by studying the transport properties of the JJs as a function of $V_{BG}$. This gate dependence arises from the electrostatic modulation of carrier density in the graphene channel, which directly affects $R_N$. Unlike conventional metallic proximity-JJs, where $R_N$ is largely set by material properties and geometry, in GJJs, the electrostatic tuning of $V_{BG}$ offers *in-situ* control via $V_{BG}$. By adjusting $V_{BG}$, the system transitions between different transport regimes—from electron-doped ($V_{BG} > V_{CNP}$) to hole-doped ($V_{BG} > V_{CNP}$), passing through the CNP ($V_{BG} = V_{CNP}$), where $R_N$ reaches its maximum. This tunability is a hallmark of 2-dimensional materials, which directly influence superconducting proximity effects. As such, Fig. 2c presents the typical $V_{BG}$ dependence of $R_N$ for a GJJ at various $L$. $R_N$ shows minimal dependence on $L$ either in highly electron- and hole-doped regions, indicating the ballistic transport in the graphene. The Fabry–Perot oscillations observed in $R_N$ at $V_{BG} < V_{CNP}$ are attributed to the quantum interference of charge carriers in the *n*–*p*–*n* heterojunction formed owing to the work function difference between the superconducting electrode and graphene [55], further confirming the ballistic transport in the graphene. We determine the contact transparency $\tau = R_Q/R_N$ as shown in the inset of Fig. 2c. Herein, $R_Q = (h/4e^2)(1/N)$ is the junction resistance expected in a ballistic limit with perfect contact transparency (black-dashed line in Fig. 2c), $h$ is Plank's constant, e is the electron charge, and $N = 2W/\lambda_F$ is the number of carrier propagation modes for a channel width ($W$) at the Fermi wavelength ($\lambda_F$). Figure 2d shows the typical $I_b$–$V$ characteristics of a GJJ, where voltage abruptly changes from the supercurrent state to the resistive state at $I_b = I_c$. As the temperature increases, $I_c$ decreases and the voltage change becomes smeared owing to the thermal fluctuations. Herein, we define $I_c$ as $I_b$ at which $V$ exceeds a threshold voltage of 1.8 µV, slightly larger than the voltage noise floor of the digital multimeter.

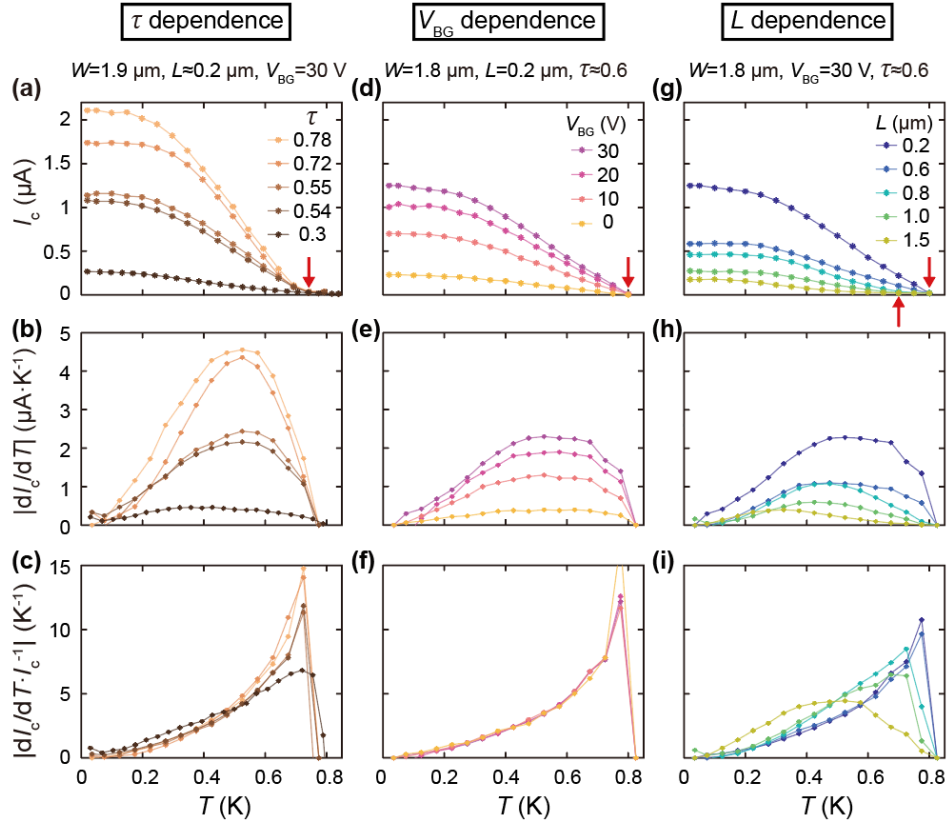

FIG 3. Junction parameter dependence in short ballistic GJJs. (a)–(c) Variations in contact transparency ($\tau$); temperature ($T$) dependencies of (a) critical current ($I_c$) for aluminum (Al) GJJs, (b) numerically calculated $|dI_c/dT|$ and (c) numerically calculated $|dI_c/dT \cdot I_c^{-1}|$ values. (d)–(f) Variations in back-gate voltage ($V_{BG}$); $T$ dependencies of (d) $I_c$, (e) $|dI_c/dT|$, and (f) $|dI_c/dT \cdot I_c^{-1}|$. (g)–(i) Variations in the channel length ($L$); $T$ dependencies of (g) $I_c$,

(h) $|dI_c/dT|$, and (i) $|dI_c/dT \cdot I_c^{-1}|$. The critical temperatures of the Josephson junction ($T_c^*$) in each case are indicated by the red arrows.

Given that the temperature dependence of $I_c$ is monotonic, a larger $|dI_c/dT|$ value is typically expected to correlate with a higher $I_c(T=0)$ and a lower critical temperature $T_c$ of the superconducting electrodes. Thus, our strategy was to shorten $L$ to increase $I_c$ and use superconducting materials with a smaller superconducting gap ($\Delta_0$), which would lead to a longer coherence length $\xi = \hbar v_F / \Delta_0$ of proximity JJ, leading to the short junction limit where $L < \xi$. Herein, $v_F$ is the Fermi velocity of charge carriers in the weak link, i.e. $10^6$ m/s for graphene. We first focus on Al as a superconducting material with a small value of $\Delta_0 \sim 0.13$ meV, and address how the parameters $\tau$, $V_{BG}$, and $L$ influence the values of $|dI_c/dT|$ and $|dI_c/dT \cdot I_c^{-1}|$. Unless the junction width is so small that disordered edges of graphene play an important role, $I_c$ is proportional to $W$; hence, $|dI_c/dT|$ should also be proportional to $W$. Note that we chose Al rather than Ti, which has a smaller $\Delta_0$ value, because the yield of getting Josephson coupling with Ti is lower than that with Al (see Methods in Supplementary Information for more details on Ti-GJJ fabrication). As the dependence on $W$ is straightforward, we did not explore it further. Figure 3a shows the temperature dependence of critical current $I_c(T)$ for Al-GJJs at different $\tau$ values. The $I_c(T)$ curves are monotonically convex upward ($d^2 I_c / dT^2 < 0$), indicating a short junction behavior as predicted by the Kulik–Omel'yanchuk (KO) theory [56]. We observe that $I_c$ increases as a function of $\tau$ while the $I_c(T)$ curves exhibit the same critical temperature of the Josephson junction ($T_c^*$), defined as the temperature above which $I_c$ falls to 3 % of its maximum value. Consequently, the value of $|dI_c/dT|$ at a given temperature increases as $\tau$ increases (Fig. 3b), in contrast to the nearly overlapping $|dI_c/dT \cdot I_c^{-1}|$ for $T < T_c^*/2$ regardless of the $\tau$ value (Fig. 3c). Figure 3d illustrates the impact of carrier dependence on $I_c(T)$ for an Al-GJJ with $\tau \approx 0.6$. We only considered the electron-doped regime ($V_{BG} > V_{CNP} = -10.4$ V) because the transport in the hole-doped regime was largely affected by the formation of p-n junctions near contacts. In the short junction limit, $I_c$ was expected to be inversely proportional to $R_N$ because the $I_c R_N$ product was solely determined by the $\Delta_0$ value at fixed $\tau$ and was not dependent on $V_{BG}$ (see Supplementary Fig. 2). All the $I_c(T)$ curves exhibit the same $T_c^*$ as KO theory predicts [56]. Consequently, $|dI_c/dT|$ increases as $V_{BG}$ increases (Fig. 3e). This contrasts to the overlapping $|dI_c/dT \cdot I_c^{-1}|$ curves regardless of the value of $V_{BG}$ at the same $T_c^*$ (Fig. 3f), which suggests $T_c^*$ would be a key parameter for determining $|dI_c/dT \cdot I_c^{-1}|$, as discussed later.

Figure 3g displays the $I_c(T)$ curves of Al-GJJs at various $L$. As $L$ decreases, the Josephson coupling strengthens, leading to an increase in $I_c$. All $I_c(T)$ curves exhibit a similar $T_c^* \sim 0.8$ K except for GJJ with $L = 1.5$ μm, which indicates the deviation from the short junction limit. The $|dI_c/dT|$ value increases as $L$ decreases (Fig. 3h). Moreover, the $|dI_c/dT \cdot I_c^{-1}|$ curves remain similar for $L = 0.2$ μm and 0.6 μm as they are in the short-junction limit (Fig. 3i). However, a deviation begins to occur at longer $L$ values as GJJ transits to the long junction limit; this will be discussed in more detail in Fig. 4.

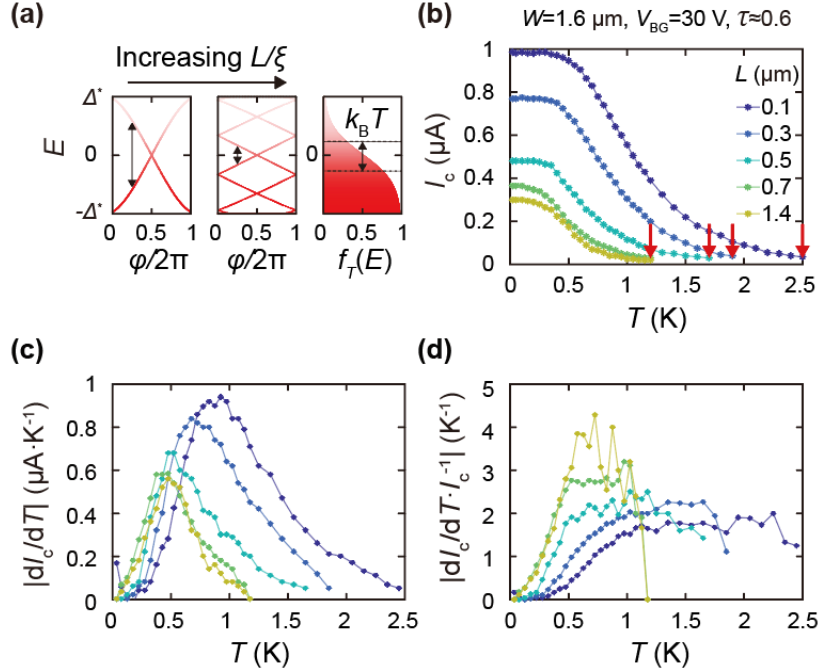

FIG 4. Junction parameter dependence in long ballistic GJJs. (a) Schematics of the Andreev bound states (ABS) from the short to long junction limit. The spectra of the ABS correspond to (left) $L/\xi = 0.25$ and (middle) $L/\xi = 3.7$. (Right) Fermi–Dirac distribution $f_T(E)$ with the thermal energy $k_B T = \Delta^*/2$. $L/\xi$ denotes the ratio between the junction length ($L$) and the superconducting coherence length ($\xi$), and $\Delta^*$ represents the proximity-induced superconducting gap in graphene. (b) Temperature ($T$) dependencies of the Josephson critical current $I_c$ in NbN-GJJs at various $L$ values. The critical JJ temperature ($T_c^*$) is indicated by the red arrows. (c, d) Numerically calculated (c) $|dI_c/dT|$ and (d) $|dI_c/dT \cdot I_c^{-1}|$ values from the $I_c(T)$ curves in (b).

We now focus on NbN-GJJs in the long junction limit and examine how $L$ affects the values of $|dI_c/dT|$ and $|dI_c/dT \cdot I_c^{-1}|$. NbN has a large value $\Delta_0 \sim 1.74$ meV and thus yields a shorter $\xi \sim 380$ nm than Al, more readily to achieve the long junction limit ($L > \xi$). To understand the temperature dependence of $I_c$, we begin by considering the occupation of the ABS bands, which follows the Fermi–Dirac distribution function $f_T(E) = 1/(e^{E/k_B T} + 1)$. The n-th ABS bands $E_n(\varphi)$ depend on the superconducting phase difference $\varphi$ between the superconducting electrodes. Specifically, total Josephson current can then be expressed as $I_J(T, \varphi) = \sum_n I_n(\varphi) f_T(E_n)$ where the contribution from each ABS band is given by $I_n(\varphi) = (2e/\hbar)\partial E_n(\varphi)/\partial \varphi$ with $\hbar = h/2\pi$. In a longer JJ, dynamical phase accumulation increases, leading to a denser spectrum of the ABS within $2\Delta^*$, as illustrated in Fig. 4a. Therefore, an increase in $L/\xi$ results in a reduction of the energy gap between the upper and lower ABS bands (indicated by black arrows) relative to the thermal energy $k_B T$, where $k_B$ is the Boltzmann constant. Consequently, longer junctions become more sensitive to temperature changes that leads to a larger $|dI_c/dT|$. The $I_c(T)$ curves of NbN-GJJs exhibit a convex downward behavior ($d^2I_c/dT^2 > 0$) at high temperatures (Fig. 4b), indicative of long junction behavior as predicted by theory [57]. In the case of the long junction limit, $I_c(T)$ follows the relationship $I_c(T) \propto \exp(-\frac{k_B L}{\hbar v_F}T)$ at high temperatures [58]. As $L$ increases, $I_c$ decreases as Josephson coupling weakens, which in turn lowers $|dI_c/dT|$. At the same time, $T_c^*$ also decreases, unlike the case of a short junction limit where $T_c^*$ is unchanged. This decrease of $T_c^*$ increases $|dI_c/dT|$. The competition of these effects on $|dI_c/dT|$ with increasing $L$ is highlighted in Fig. 4c.

For the inductive readout scheme, $|dI_c/dT \cdot I_c^{-1}|$ increases as a function of $L$ (Fig. 4d), suggesting that small $T_c^*$ values are crucial for maximizing $|dI_c/dT \cdot I_c^{-1}|$. A longer $L$ could be advantageous for achieving a smaller $T_c^*$, which in turn may enhance the thermal sensing performance. However, as $L$ increases, the area or the volume of weak link material also expands, which increases its heat capacity and consequently reduces the electron temperature rise from a quantized excitation, thereby degrading detector performance.

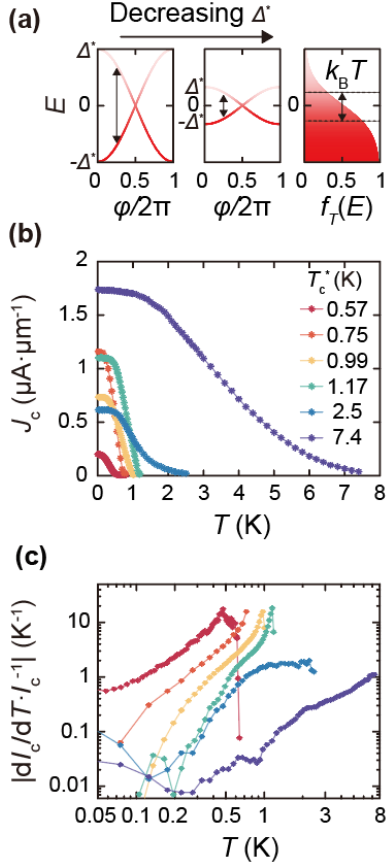

FIG 5. JJ's critical temperature ($T_c^*$) dependence in GJJs. (a) Schematics of the ABS spectra for (left) a large $\Delta^*$, (middle) small $\Delta^*$, and (right) the Fermi–Dirac distribution $f_T(E)$ with thermal energy $k_B T = \Delta^*/2$. (b) Temperature ($T$) dependence of the Josephson critical current density $J_c = I_c/W$, and (c) numerically calculated $|dI_c/dT \cdot I_c^{-1}|$ values for GJJs made of various superconducting materials. Colors indicate the different fabrication recipes: Ta/Ti/Au (red), Ti/Al/Au (orange), Ti/Al(thin) (yellow), Ti/Al(thick) (green), Ti/Nb/NbN (blue), and MoRe (purple).

An alternative way to tune $T_c^*$ without affecting graphene's heat capacity is to engineer the superconducting gap through the selection of adhesion layers (Ti, Ta, Ti/Nb), superconducting materials (Ti, Al, NbN, MoRe), and capping layers (Au) [59] and their respective thicknesses. $T_c^*$ is directly proportional to $\Delta^*$, which encapsulate all junction-related parameters, including $L$, $V_{BG}$, $\tau$, as well as the properties of the adhesion layers and superconducting materials. The schematics of the ABS in the short junction limit, illustrated in Fig. 5a, show that a smaller $\Delta^*$ results in a narrower energy gap between upper and lower ABS bands, leading to a faster $I_c$ decay at increasing temperatures. Figure 5b displays the temperature dependence of the Josephson critical current density ($J_c = I_c/W$) for GJJs fabricated with various recipes and junction parameters (see Table I for device information). While direct comparison of $|dI_c/dT|$ is not feasible due to differences in $L$, $V_{BG}$, and $\tau$ across devices, $|dI_c/dT \cdot I_c^{-1}|$ can be compared as done in Fig. 3. Figure 5c presents a double logarithmic plot of the temperature dependence on $|dI_c/dT \cdot I_c^{-1}|$, demonstrating the impact of $T_c^*$. As indicated, smaller $T_c^*$ values yield larger $|dI_c/dT \cdot I_c^{-1}|$ values. For instance, a Ti-GJJ with the lowest $T_c^* = 0.57$ K exhibits the largest $|dI_c/dT \cdot I_c^{-1}|$ value. $\Delta^*$ can be as large as $\Delta_0$ in a short junction limit with perfect contact transparency and perfect conduction in graphene. By contrast, in a long junction limit, $\Delta^*$ is smaller than $\Delta_0$ and decreases as a function of $L$, expressed as $\Delta^* \propto \hbar v_F/L$ [55]. Therefore, $\Delta^*$ needs to be considered to optimize the thermal detector performance instead of $\Delta_0$. Notably, when small-gap superconducting materials are chosen to achieve a small $\Delta^*$, the junction width needs to be sufficiently increased to preserve the Josephson coupling. However, this process will enlarge the graphene area, which in turn reduces the electron temperature rise. Therefore, careful optimization of the junction parameters is required.

TABLE I. Summary of junction parameters of graphene Josephson junction devices presented in Fig. 5. Numbers in parentheses in the "Recipe" column represent the thickness of each film in nanometers.

| Recipe | $T_c^*$ (K) | $\xi$ (μm) | $L$ (μm) | $W$ (μm) | $V_{BG}$ (V) | $\tau$ | $I_c$ (20 mK) (μA) | $R_N$ (Ω) | $I_c R_N$ (μeV) |
|---|---|---|---|---|---|---|---|---|---|
| Ta (10) / **Ti** (60) / Au (5) | 0.57 | 7.2 | 0.2 | 5.3 | 30 | 0.3 | 1.08 | 43.3 | 46.8 |
| Ti (6) / **Al** (60) / Au (5) | 0.75 | 5.1 | 0.2 | 1.8 | 30 | 0.78 | 2.11 | 42.0 | 88.6 |
| Ti (6) / **Al** (70) | 0.99 | 4.4 | 0.3 | 2.9 | 20 | 0.53 | 2.11 | 64.9 | 137 |
| Ti (6) / **Al** (200) | 1.17 | 3.6 | 0.3 | 2.9 | 45 | 0.42 | 3.17 | 48.2 | 153 |
| Ti (6) / Nb (5) / **NbN** (50) | 2.5 | 0.38 | 0.1 | 1.6 | 30 | 0.58 | 0.985 | 72.9 | 71.8 |
| **MoRe** (50) | 7.4 | 0.46 | 0.2 | 1.8 | 30 | 0.27 | 3.13 | 133 | 419 |

## IV. CONCLUSIONS

We presented our efforts to optimize the junction parameters for thermal sensing in two different operation schemes—the threshold and the inductive. These schemes respectively require the maximization of $|dI_c/dT_e|$ and its normalized counterpart $|dI_c/dT_e \cdot I_c^{-1}|$. In the short junction limit, the shorter $L$, higher $V_{BG}$, and greater $\tau$ values result in larger $|dI_c/dT|$ values—favorable for threshold measurements. However, the value of $|dI_c/dT \cdot I_c^{-1}|$ for the inductive scheme remains unaffected by the changes in $L$, $V_{BG}$, or $\tau$ as $T_c^*$ remains constant. In the long junction limit, a longer $L$ increases $|dI_c/dT \cdot I_c^{-1}|$ owing to a smaller $T_c^*$, but also enlarges the size of the weak link, increasing its heat capacity and reducing the electron temperature rise induced by photon absorptions.

From the threshold scheme, we achieved $|dJ_c/dT| \approx 0.2\ \mu\text{A}\ \text{K}^{-1}\mu\text{m}^{-1}$ at 0.1 K for an Al-GJJ with $W = 1.8\ \mu\text{m}$, $L = 0.2\ \mu\text{m}$, $\tau = 0.78$, and $V_{BG} = 20\ \text{V}$ corresponding to a carrier density of approximately $1.7 \times 10^{16}\ \text{m}^{-2}$, implying that an Al-GJJ with $W = 5\ \mu\text{m}$ and $L = 0.2\ \mu\text{m}$ will achieve $|dI_c/dT| = 1\ \mu\text{A/K}$ with an electronic graphene heat capacity of $\sim 6\ k_B$ [20]. Based on a previous study [20], with an appropriate setting of $I_b/I_c = 0.91$, this value of $|dI_c/dT|$ is sufficient for detecting a single photon at 26 GHz with a quantum efficiency > 0.99 and a dark count probability < 0.07. For the inductive scheme, we demonstrated a Ti-GJJ value with $T_c^* = 0.57\ \text{K}$ and $|dI_c/dT \cdot I_c^{-1}| = 0.6\ \text{K}^{-1}$ at 50 mK, achieved by engineering the $\Delta^*$ value based on various fabrication recipes, which included the selection of appropriate adhesion layers, superconducting materials, capping layers, and their respective thicknesses.

These results establish a systematic approach to optimizing the thermal sensitivity of proximity JJs by engineering ABSs. By identifying the key role of the proximity-induced superconducting gap and critical device parameters—such as channel length, transparency, and carrier density—we provide a versatile strategy for enhancing thermal response of proximity JJs. More broadly, this work advances the study of ABS-mediated transport in proximity JJs, while informing the design of the next-generation ultra-sensitive JJ-based sensors.

## ACKNOWLEDGMENTS

This work was supported by the National Research Foundation (NRF) Grants (Nos. RS-2022-NR068223, RS-2024-00393599, RS-2024-00442710, RS-2024-00444725) and the ITRC program (IITP-2025-RS-2022-00164799) funded by the Ministry of Science and ICT, Samsung Science and Technology Foundation (Nos. SSTF-BA2401-03 and SSTF-BA2101-06), and Samsung Electronics Co., Ltd. (IO201207-07801-01). S.J. was supported by a Basic Science Program through the NRF funded by the Ministry of Education (No. 2022R1A6A3A01086903). E.G.A. was supported by the Army Research Office MURI (Ab-Initio Solid-State Quantum Materials) Grant no. W911NF-18- 1-043. K.W. and T.T. acknowledge support from the JSPS KAKENHI (Grant Numbers 21H05233 and 23H02052) and World Premier International Research Center Initiative (WPI), MEXT, Japan. E.G.A. was supported by the Army Research Office MURI (Ab-Initio Solid-State Quantum Materials) Grant no. W911NF-18-1-043. B.H. was supported by an appointment to the Intelligence Community Postdoctoral Research Fellowship Program at the Massachusetts Institute of Technology, administered by Oak Ridge Institute for Science and Education through an interagency agreement between the U.S. Department of Energy and the Office of the Director of National Intelligence.

# Supplementary Information for

**Engineering Andreev Bound States for Thermal Sensing in Proximity Josephson Junctions**

Woochan Jung[1], Ethan G Arnault[2], Bevin Huang[3], Jinho Park[1,4], Seong Jang[1], Kenji Watanabe[5], Takashi Taniguchi[6], Dirk Englund[7], Kin Chung Fong[8,9,10,*], Gil-Ho Lee[1,*]

[1]Physics Department, Pohang University of Science and Technology, Pohang 37673, South Korea
[2]Research Laboratory of Electronics, Massachusetts Institute of Technology, Cambridge, MA 02139, USA
[3]Intelligence Community Postdoctoral Research Fellowship Program, Massachusetts Institute of Technology, Cambridge, MA 02139, USA
[4]Department of Mechanical Engineering, Columbia University, New York, NY 10027, USA
[5]Research Center for Functional Materials, National Institute for Materials Science, Tsukuba, Japan
[6]International Center for Materials Nanoarchitectonics, National Institute for Materials Science, Tsukuba, Japan
[7]Department of Electrical Engineering and Computer Science, Massachusetts Institute of Technology, Cambridge, MA 02139, USA
[8]Department of Electrical and Computer Engineering, Northeastern University, Boston, MA 02115, USA
[9]Department of Physics, Northeastern University, Boston, MA 02115, USA
[10]Quantum Materials and Sensing Institute, Burlington, MA 01803, USA

*Corresponding authors: k.fong@northeastern.edu (K.C.F.), lghman@postech.ac.kr (G.-H.L.)

**The file includes:**

- Methods
- Figs. S1 to S2

# Methods

1. General GJJ fabrications

First, graphene and hexagonal boron nitride (hBN) flakes were exfoliated on separate silicon oxide wafers. We employed a dry transfer method [Ref] to encapsulate the graphene with hBN flakes. This encapsulated stack was then placed on silicon oxide wafers (280 nm $SiO_2$) and annealed under vacuum $(< 1 \times 10^{-3}$ Torr$)$ at 500 °C for 1 hour. Subsequently, the stack was shaped into a rectangle using reactive ion etching with $CF_4$ and $O_2$ plasmas, and the electrodes were patterned using a standard electron-beam lithography. Superconducting edge contacts were made to the freshly etched edges of graphene by depositing materials via electron-beam evaporation or DC sputtering.

2. Aluminum based Graphene Josephson junctions (Al-GJJs)

We fabricated several Al-GJJs to explore different contact transparency $(\tau)$ parameters, achieved by adjusting electron-beam lithography dose conditions, Methyl Isobutyl Ketone (MIBK) developer times, plasma etching conditions, and the base pressure of the electron-beam deposition chamber.

After shaping the stack into a rectangle, we patterned it using standard e-beam lithography with a 950 PMMA C4 mask. **The patterning process included:**

- Oxygen plasma etching: The sample was cleaned in $O_2$ plasma for 10 seconds (conditions: 4 sccm, 50 W forward power, 17 W reflected power, and 200 mTorr).
- $CF_4$ plasma etching: To etch the top hBN layer, we applied $CF_4$ plasma under two sets of conditions: (condition1: 40 sccm, 50 W forward power, 2 W reflected power, and 1070 mTorr) or (condition2: 4 sccm, 5 W forward power, 0 W reflected power, and 90 mTorr).
- Second oxygen plasma etching: Another round of $O_2$ plasma etching for 20 seconds was performed under the initial conditions to etch the graphene.
- Final $CF_4$ plasma etching: A repeat of the $CF_4$ plasma etching was conducted to etch the bottom hBN layer using the previously mentioned conditions.

**Edge contacts:** Superconducting edge contacts to the graphene layer were established by in situ electron-beam deposition, using a Ti adhesion layer followed by an Al superconducting layer, with an optional Au capping layer added. Al was deposited using various recipes to fine-tune the critical temperature $(T_c)$ of the electrodes. The specific recipes used are summarized in Table 1 of the main text and described in detail as follows:

(1) Ti/Al/Au: A 60 nm thick Al film with a 6 nm Ti adhesion layer and a 5 nm Au capping layer was used to engineer the superconducting gap. The Au acts as a normal-metal layer, inducing an inverse proximity effect.
(2) Ti/Al (thin): A 70 nm thick Al film with a 6 nm Ti adhesion layer was used. The thickness of the Al film modulates its $T_c$, with thinner films typically having a lower $T_c$ due to increased surface scattering, quantum size effects, and higher levels of disorder.
(3) Ti/Al (thick): A 200 nm thick Al film with a 6 nm Ti adhesion layer was used to explore the effects of a thicker superconducting layer.

3. Titanium based Graphene Josephson junctions (Ti-GJJs)

For Ti-GJJs, there was an absence of Josephson coupling when using Ti as an adhesion layer. The likely issue is that due to the high melting temperature of Ti, continuous outgassing from the electron beam resists occurs during Ti deposition, which is suspected of contaminating the superconductivity. This problem can be resolved by using tantalum (Ta) as an adhesion layer, although further optimization of the fabrication process is needed to improve yield.

In our experiment, we used 60 nm thick films of Ti with a 10 nm adhesion layer of Ta. Additionally, a 5 nm Au capping layer was applied to protect the Ti from oxidation. Initially, the 10 nm Ta layer was deposited by a sputtering process using a DC plasma at 300 W in an Ar atmosphere at 3.4 mTorr. Without breaking vacuum, a 60 nm Ti layer was then deposited by electron-beam evaporation at a base pressure in the low $10^{-7}$ Torr range, followed by the deposition of a 5 nm of Au capping layer on the Ti.

4. Molybdenum-rhenium based Graphene Josephson junctions (MoRe-GJJs)

Superconducting edge contacts were made to the freshly etched edges of graphene by sputtering MoRe materials. A 50 nm layer of MoRe was deposited by a sputtering process using a DC plasma at 300 W in an Ar atmosphere at 3.4 mTorr.

5. Niobium nitride-based Graphene Josephson junctions (NbN-GJJs)

On the edge of graphene freshly revealed by plasma etching, a 6 nm Ti was deposited by electron-beam evaporation, followed by sputtering a 5 nm Nb layer and a 50 nm NbN superconducting layer, all without breaking vacuum. For NbN film, reactive sputtering was employed in an Ar/$N_2$ atmosphere (50 sccm:9 sccm) at 3.4 mTorr. A thin Nb film was deposited prior to the NbN film formation to protect the initial Ti layer from the reactive N radicals during NbN film sputtering.

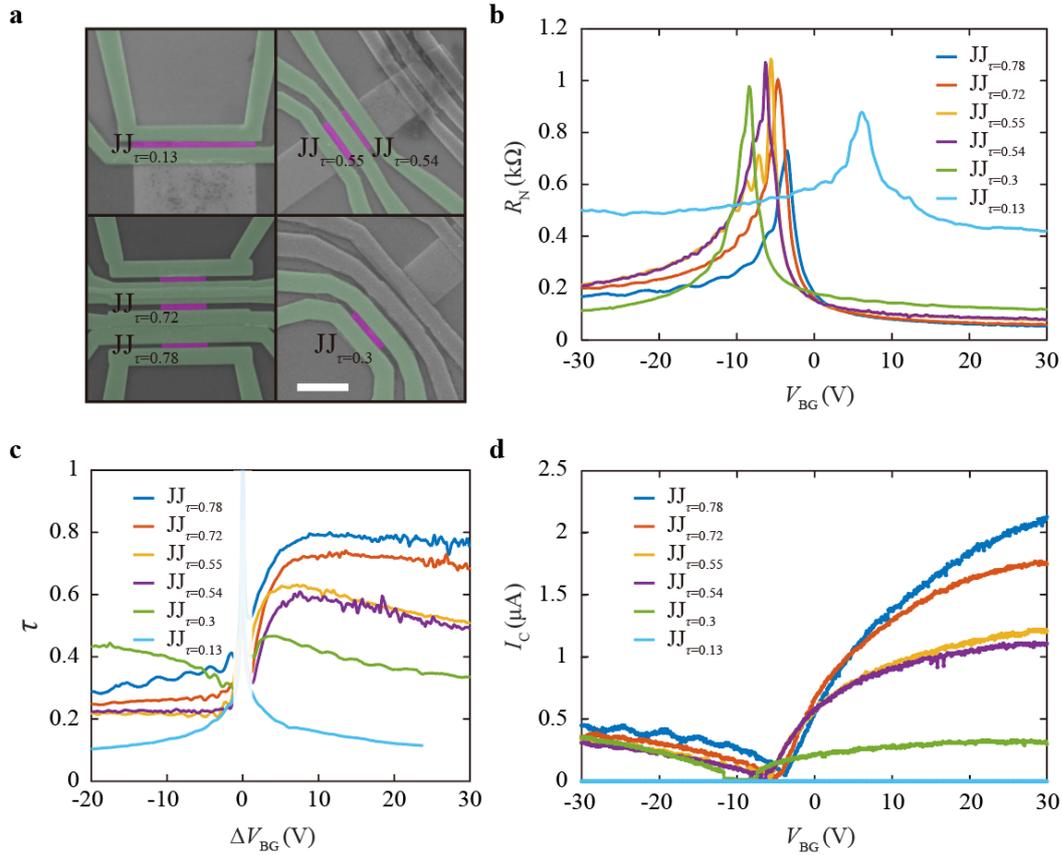

**Supplementary Fig. 1 | Transport Characterization of Aluminum-Graphene Josephson Junctions. a,** False-colored Scanning Electron Microscope (SEM) image of Graphene Josephson junctions (GJJs). Aluminum electrodes (green) make ohmic contacts with graphene (magenta). Each Josephson junction (JJ) is identified by the transparency ($\tau$) value it has at backgate voltage ($V_{BG}$) of 30 V. The scale bar represents 2 μm. **b,** Normal-state junction resistance ($R_N$) for each GJJ measured at 4.2 K across varying $V_{BG}$. **c,** Each $\tau$ plotted as a function of $\Delta V_{BG}$, which is defined as the voltage deviation from the charge neutrality point. **d,** $V_{BG}$ dependence of the critical current ($I_c$) for each junction measured at 20 mK.

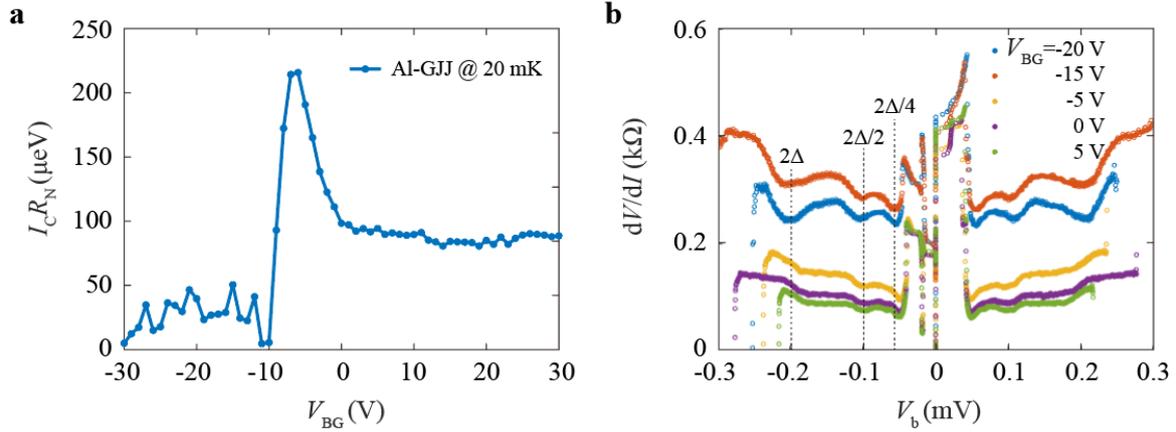

**Supplementary Fig. 2 | $I_cR_N$ product of Aluminum-Graphene Josephson Junction. a,** $I_cR_N$ product for the junction, corresponding to the data in Figure 3d of the main text. This product is independent of the $V_{BG}$ when in positive values. **b,** Bias spectroscopy showing dips in the differential resistance of the junction due to multiple Andreev reflections. These dips are utilized to estimate $\Delta_0$. The value of $eI_cR_N/\Delta_0$ reaches approximately 1 for positive gate voltages well beyond the Dirac point.